\documentclass{elsart}
\usepackage{graphicx,amssymb,amsmath,color}
\def\cal{\mathcal}
\def\e{{\rm e}}
\def\l{\left(}
\def\r{\right)}

\begin{document}
\begin{frontmatter}
\title{
Towards event-by-event studies of the ultrahigh-energy cosmic-ray
composition}

\author{D.\ S.\ Gorbunov$^a$,}
\author{G.\ I.\ Rubtsov$^{a,b}$ and}
\author{S.\ V.\ Troitsky}$^a$

\address{$^a$Institute for Nuclear Research of the Russian Academy of
Sciences,\\
60th October Anniversary Prospect 7a, 117312, Moscow, Russia\\
$^b$ Moscow State University, Vorobiovy Gory, Moscow, Russia
}
\ead{st@ms2.inr.ac.ru}
\begin{abstract}
We suggest a method which improves the precision
of studies of the primary composition of
ultra-high-energy cosmic rays. Two principal ingredients of the method are
(1)~comparison of the observed and simulated parameters for
{\em individual}
showers, without averaging over arrival directions and (2)~event-by-event
selection of simulated showers by the physical observables and not by
the reconstructed primary parameters. A detailed description of the
algorithm is presented and illustrated by several examples.
\end{abstract}
\date{}

\end{frontmatter}

\section{Introduction}
\label{sec:intro}
The determination of the primary composition of cosmic rays with energies
higher than $\sim 10^{17}$~eV is a real challenge. The lack of knowledge
of the types of primary ultra-high-energy particles which induce extensive
air showers makes it difficult to study their origin and in some cases
even to determine their energy spectrum. More precise and less
model-dependent determination of the cosmic-ray primary composition,
especially in the highest-energy domain, is one of the most important
tasks in contemporary astroparticle physics (for a review and discussions
see e.g.\ Ref.~\cite{Watson}).

Ultra-high-energy cosmic rays are observed through air showers, and to extract
information about the primary particle, selected observables of real
air showers are compared to those of  simulated ones for different
primaries. Because of large shower-to-shower fluctuations, one cannot
determine the primary-particle type of an individual air shower. That
is why traditional approaches to the composition studies are based on
the determination of average characteristics of a large sample of
cosmic-ray events. This approach has obvious advantages: for a
homogeneous composition, averaging smoothens out the fluctuations, and
large statistics results in higher precision. Furthermore, the
computational time required to perform reliable simulations of an
observable averaged over a large sample is much smaller than that
necessary for detailed simulations of all events in the data
set. However, for mixed composition, averaging might become a problem
because fluctuations of the discriminating observables over their
central values are larger when air showers from different arrival
directions are combined in one sample. Different zenith angles
correspond to different atmospheric depths, and air showers detected
by surface arrays have different stages of development and hence may
have quite different observable parameters even for the same energies
and primaries.  Moreover, the geomagnetic field introduces the
azimuthal dependence for photon~\cite{gamma-gMF} and very inclined
hadron~\cite{inclined-gMF} showers. Even fluorescent detectors, which
observe the shower development on its way through the atmosphere,
respond to showers from different arrival directions in different
ways, notably with different accuracies.  For gamma-ray primaries at
$E\gtrsim 5\cdot10^{19}$~eV, the entire shower development is
direction-dependent.

We suggest to improve considerably the precision of the composition
studies by performing individual simulations for each observed
high-energy event. In the case of a large number of events in a data
sample,
averaging in bins of arrival directions may be used.
If one is not interested in the study of possible gamma-ray
primaries with $E\gtrsim 5\cdot10^{19}$~eV,
binning in zenith angle only is often sufficient.

This method, in its simplest form with one observable, has been
successfully implemented to obtain a limit on the gamma-ray fraction in
the primary flux at $E>10^{20}$~eV using the data of the AGASA and Yakutsk
experiments~\cite{gamma-limit}. Event-by-event simulations (without
selection by reconstructed parameters) were previously used for the
composition studies in Refs.~\cite{Fly-s-Eye,Homola,Fly-s-Eye-Homola}.

The rest of the paper is organized as follows. In
Sec.~\ref{sec:general}, we present the sketch of the event-by-event
approach to the composition studies, discuss the choice of the
air-shower observables (Sec.~\ref{sec:EandC}), give a general idea of
how to constrain the probable primary-particle type of an individual
air shower (Sec.~\ref{sec:general-event}) and how to use these
shower-by-shower constraints to gain information about the chemical
composition of the primary cosmic-ray flux using a sample of events
(Sec.~\ref{sec:general-ensemble}). Sec.~\ref{sec:integrals} contains
the detailed description of the procedure outlined in
Sec.~\ref{sec:general} and ready-to-use formulae implementing this
procedure.  Sec.~\ref{sec:example} presents several examples which
illustrate the method by the analysis of small (and hence
statistically insignificant) samples of events. There, we consider
only one composition-related observable, the muon density, and use
samples of highest-energy AGASA and Yakutsk events as examples. In
Sec.~\ref{sec:example-event}, we analyse in detail a single event (the
highest energy air shower reported by AGASA). The procedure for
estimating the limit on the fraction of particular primaries in a
given energy range is illustrated in
Sec.~\ref{sec:example-gamma-limit} with the sample of six AGASA
showers with reported energies higher than $10^{20}$~eV and known muon
content, while the algorithm to determine the best-fit composition
assuming two possible primaries is illustrated in
Sec.~\ref{sec:example-p-Fe} with a sample of four events with reported
energies above $1.5\cdot 10^{20}$~eV detected by the AGASA and Yakutsk
experiments. Both samples are small and the analysis of
Sec.~\ref{sec:example} serves for illustrative purposes only. We
briefly summarize and discuss novelties of our method in
Sec.~\ref{sec:concl}.  Some of our notations are summarized in
Appendix~\ref{app:notations}.  Appendices~\ref{app:events} and
\ref{app:reconstruction} contain technical information related to the
examples presented in Sec.~\ref{sec:example}.

\section{Generalities}
\label{sec:general}
In this section, we sketch the main elements of our approach to
study individual air-shower events and their ensembles. An operational
algorithm is given in Sec.~\ref{sec:integrals} and illustrated in
Sec.~\ref{sec:example}.

\subsection{Two classes of observables}
\label{sec:EandC}
On the basis of detector readings, several air-shower observables can
be determined experimentally; many of them are not independent of each
other. For our purposes, we separate them into two groups which we will
treat differently.

\subsubsection{E-observables}
This class contains the parameters related to the energy estimation
and the arrival direction. For ground arrays, E-observables usually
include the signal density at a given distance from the shower core
(known as $S(600)$ or $S(1000)$); for fluorescent telescopes the
E-observable is the total amount of the fluorescent light (corrected
for atmospheric conditions).

The arrival direction may be fixed because the pointing accuracy at
high energies is often relatively high and variations of the arrival
direction within the error bars have a negligible, compared to
shower-to-shower fluctuations, effect on both
energy estimation and measurement of the composition-related parameters.
In the case of poor angular resolution, for
example for fluorescent detectors in the monocular mode, the arrival
direction should be treated on the same footing as other E-parameters.

In what follows, we ignore, for simplicity, the errors in
the determination of the arrival direction and consider a single E-observable
--- the reconstructed energy of a shower, $E_{\rm rec}$. For a fixed
arrival direction, $E_{\rm rec}$ is in one-to-one correspondence with the
energy estimator used by a given experiment.
 
We note that E-observables are primary-dependent but this dependence is
naturally accounted for in the method, see Sec.~\ref{sec:general-event}.

\subsubsection{C-observables}

We call the parameters used to distinguish various types of primary
particles C-observables. Depending on the experiment, they may include
muon density or muon richness, the slope of the lateral distribution
function of the total signal or of muons only, shower front curvature,
rise time etc. For a more efficient separation of different primaries,
simultaneous use of several such parameters is justified, and we hereafter
denote the set of these parameters as ${\bf c}=(c_1,c_2,\dots)$.

\subsection{Study of an individual event}
\label{sec:general-event}
From the sample of air showers of interest, each event is to be
studied separately. One simulates a number of artificial showers whose
E-observables are consistent with the real event and which are
initiated by different primaries.  To this end, showers with different
primary energies are simulated; for each of them, E-observables are
reconstructed and compared to their observed values for the real
shower. Simulated events are then selected for further study by
assigning weights proportional to the estimate of how well their
E-observables match the data. It is important to use the same
reconstruction procedure as in the analysis of the real data; detailed
information about the detector is needed at this point.
 
Any simplified treatment of the detector (unless quantified to be
of negligible effect) would weaken the improvements of the proposed method
over previous ones.  

For each simulated shower, C-observables are reconstructed, again by the
same method as used for processing the real data. For each type of
primary particle of interest, the distribution of C-observables of
simulated showers selected by E-observables is obtained.

Finally, the C-observables measured for the real event are compared
with these simulated distributions to determine the probabilities that
the event was initiated by various primaries, given the measured E-
and C-observables.

\subsection{Study of an ensemble of events}
\label{sec:general-ensemble}
One is usually interested in a certain range of real primary energies
$E_0$ (see Appendix~\ref{app:notations} for the summary of our
notations). We will refer to this range as $\{E\}$; it may be an
interval ($E_1<E_0<E_2$) or a half-line ($E_0>E_1$); in
general, the restrictions on the energy may be supplemented by those
on the arrival directions if a particular region of the sky is
studied.

One then selects the data set to be analyzed. While it should more
or less correspond to the energy range of interest in terms of the
reconstructed energies, the possibility of incorrect energy reconstruction
has to be taken into account and the range of reconstructed energies
should preferrably be extended. In an ideal case,
all events recorded by an experiment should enter the sample, but most of
them would contribute to the quantities of interest with almost zero
weight.

Each of the individual events in the real sample is to be analyzed as
described in Sec.~\ref{sec:general-event} but keeping track of the
original energy $E_0$. Namely, for each event and each primary
particle, two distributions are to be obtained for the C-parameters of
the simulated events: (i)~consistent with the observed one by
E-parameters and having thrown energies in the domain of interest,
$E_0 \in \{E\}$; and (ii)~consistent with the real event by
E-parameters and having thrown energies outside the domain of
interest, $E_0 \notin \{E\}$. Separate probabilities are to be
obtained for the cases (i) and (ii).
 
This is necessary because we want to constrain primary composition in the
given domain of physical energies $\{E\}$, and contamination by showers
with energies outside $\{E\}$ should be properly taken into account.  

Then, the ensemble of these probablities obtained for all real events in
the sample is subject to combinatoric analysis.
As a result,
either limits or best-fit
parameters of the chemical composition of the primary cosmic-ray flux in
the energy domain $\{E\}$ are determined.
At this last stage of the procedure one takes into account corrections
for the ``lost events''.
These are possible events with thrown
energies $E_0 \in \{E\}$ which however would escape from the sample
either because their reconstructed energies differ strongly from $E_0$, or
because of the event quality cuts.

\section{Implementation}
\label{sec:integrals}
In this section, we present a detailed algorithm which can be directly
used in the analysis of the real data.

\subsection{Study of an individual event}
\label{sec:integrals-event}
Let us consider a single real event with the reported observed
parameters $(E_{\rm obs},{\bf c}_{\rm obs})$ and the arrival direction
$(\theta, \phi)$ in horizonthal coordinates. Hereafter, $A$ denotes
possible primary-particle type (for example, $A=p$ refers to primary
protons, $A=$Fe to iron nuclei and $A=\gamma$ to photons). For each
$A$ one is interested in, one generates a library of simulated
showers which have:

\hskip 0.3cm
{\it (i)} primary particle $A$;

\hskip 0.3cm
{\it (ii)}
arrival direction $(\theta, \phi)$ (or various arrival directions
consistent with $(\theta, \phi)$ within the experimental errors, as
discussed in Sec.~\ref{sec:EandC});

\hskip 0.3cm
{\it (iii)}
energies $E_0$ in a relatively wide range around $E_{\rm obs}$.

The thrown energies $E_0$ of the simulated showers may be chosen randomly
from, e.g., the interval $(0.5 E_{\rm obs}, 5 E_{\rm obs})$ according to $1/E_0$
spectrum. This choice of the spectrum for the library enables one to control
shower-to-shower fluctuations at high energies but saves computational
time; as we will see below, it is {\bf not} the spectrum of real particles
which we assume to be realized in Nature. The interval we quote is
indicative and should be adapted for particular events, especially at
large zenith angles.

The measurement of E-parameters is subject to statistical errors which
are studied by experimental groups in detail. The probability
distribution that the primary particle which produced an actual shower
with the observed E-parameters equal to
$E_{\rm obs}$ would rather produce a shower with these parameters
equal to $E_{\rm rec}$ is denoted by $g_E(E_{\rm rec},E_{\rm
  obs})$. This function is usually determined and published by
experimental groups.
For instance, for the AGASA experiment
$g_E(E_{\rm rec},E_{\rm obs})$ is Gaussian in $\log(E_{\rm rec}/E_{\rm
obs})$ and the standard deviation of $E_{\rm rec}$ is $\sigma_E
\approx 0.25 E_{\rm obs}$ \cite{AGASAenergy}.
Hence, in a library of simulated showers to be used in the analysis of
a real shower with E-parameters equal to
$E_{\rm obs}$, we assign to each simulated shower a weight
$$
w_1=g_E(E_{\rm obs},E_{\rm rec})\;.
$$
In principle, this function may depend on the type of a primary, but
at this stage we use one and the same $g_E$ for all particles.
 
We emphasise that the function $g_E$ has nothing to do with physical
fluctuations in the shower development, nor with primary- and
direction-dependent systematics in the energy reconstruction, but
describes instead the resolution power of the installation.  

Additionally, one may be interested in studying
the energy spectrum different from the one used in the simulations of shower
library (it may happen, for instance, that the library was simulated with
the $E_0^{-\alpha _{\rm lib}}$ spectrum with $\alpha _{\rm lib}=1$ while
one is interested in say $\alpha =2$ or $\alpha =2.7$). To reproduce the
required thrown energy spectrum, an additional weight
\begin{equation}
w_2=\left( {E_0\over E_{\rm obs}}   \right) ^{\alpha _{\rm lib} - \alpha}
\label{w2}
\end{equation}
is introduced (instead of $E_{\rm obs}$ in the denominator, any other
typical energy scale may be used).

The C-parameters are also reconstructed with some statistical
errors. In exactly the same way a shower with measured C-parameters
equal to ${\bf c}$ could produce detector readings corresponding to
 ${\bf c}'$. We denote the corresponding probability distribution as
\[
g_c({\bf c}',{\bf c}).
\]
Let us enumerate the showers in the library simulated for a given real
event and for a given primary $A$ by $i=1,\dots,M$. The distribution of the
parameters $\bf c$ for the showers consistent with the real one by
E-parameters is given by
\begin{equation}
f_A({\bf c})=\frac{1}{\cal{N}}\sum\limits_{i=1}^M g_c({\bf c},{\bf
  c}_{iA})
w_{1,iA} w_{2,iA},
\label{f_A(c)}
\end{equation}
where we denoted by $w_{1,iA}$, $w_{2,iA}$ and ${\bf c}_{iA}$ the values
of the weights and the vector of C-parameters calculated for the simulated
shower number $i$ with the primary $A$, respectively. $\cal{N}$ is the
normalization factor determined by the condition
\[
\int f_{A}({\bf c})d{\bf c}=1\;.
\]

By definition, $f_A({\bf c})d{\bf c}$ is the probability to have observed
C-parameters in the region $d{\bf c}$ centered at $\bf c$ given the
primary particle $A$ and observed E-parameters. This probability assumes
that a specific shower-development model used in the simulation code is
valid.

To compare the simulated distributions $f_A({\bf c})$ with the observed
C-parameters ${\bf c}_{\rm obs}$ of the real event, one has to formulate a
conjecture about the nature of its primary particle.

One possible
question is, how bad the event is described
by the primary $A$ -- without fixing and studying other possibilities
(only the shower library for $A$ primaries is required in this case).
Given the observed values of $E_{\rm obs}$ and ${\bf c}_{\rm obs}$, the
event is consistent with the primary $A$ with the probability density
\begin{equation}
\tilde p_{A}=f_{A}({\bf c}_{\rm obs}).
\label{p_A-tilde}
\end{equation}
In the case of practical interest, when the event is unlikely being
initiated by the primary $A$, the estimate of the probability that a
given event could be initiated by the primary $A$ is given by
\begin{equation}
\label{estimate-bound}
p_{A_1}=F_{A}({\bf c}_{\rm obs})\equiv \int\limits_{f_{A}({\bf c})
\le f_{A}({\bf c}_{\rm obs})} f_{A}({\bf c}) d{\bf c}, ~~~~ k=1,2.
\end{equation}
The use of this estimate is justified if $p_{A_1}\ll 1$.

Another conjecture to be studied might be that the primary was either
$A_1$ or $A_2$. In this case, $p_{A_1}+p_{A_2}=1$ and
\begin{equation}
p_{A_k}={f_{A_k}({\bf c}_{\rm obs}) \over f_{A_1}({\bf c}_{\rm obs}) +
f_{A_2}({\bf c}_{\rm obs})}.
\label{p_A1,A2}
\end{equation}
Clearly, it may happen that the event is poorly described by both $A_1$
and $A_2$, that is $f_{A_1}\approx f_{A_2}\approx 0$ and $p_{A_{1,2}}$
are not stable. This indicates that the conjecture that the primary was
either $A_1$ or $A_2$ works poorly for this event.

In case one wishes to distinguish between several possible primaries
$A_1,\dots,A_K$, Eq.~(\ref{p_A1,A2}) should be replaced by
\[
p_{A_i}=\frac{f_{A_i}({\bf c}_{\rm obs})}{\sum_{k=1}^{K}
f_{A_k}({\bf c}_{\rm obs})}\;.
\]

\subsection{Study of the ensemble of events}
\label{sec:integrals-ensemble}
Let us fix the physical energy range $\{E\}$ and consider a sample of $N$
observed events enumerated by $j=1,\dots,N$, selected as described in
Sec.~\ref{sec:general-ensemble}.

\subsubsection{One primary type}
\label{sec:ensemble-1-primary}
To constrain the fraction of the primaries $A$ in the total flux of
cosmic-ray particles within the energy domain $\{E\}$, one simulates
libraries of $A$-induced showers described in
Sec.~\ref{sec:integrals-event} for each observed event in the sample. When
determining the distribution $f_A({\bf c})$, one has to distinguish its
two constituents,
\begin{equation}
f_A({\bf c})=f^{(+)}_A({\bf c})+f^{(-)}_A({\bf c}),
\label{f+-}
\end{equation}
where $f^{(+)}_A({\bf c})$ is the distribution built with only those showers
from the library whose thrown energies $E_0$ belong to the domain $\{E\}$;
the rest of showers contribute to $f^{(-)}_A({\bf c})$. Consequently, one
obtains for each observed event $j$, two probabilities:
\begin{itemize}
 \item
$p^{(+)j}_A$, the probability that the event $j$ was initiated by the
primary $A$ with energy $E_0 \in \{E\}$; in the case when the event
$j$ is unlikely to be induced by primary $A$, $p^{(+)j}_A$ is given by
Eq.~\eqref{estimate-bound} with $f_A({\bf c})= f^{(+)}_A({\bf c})$,
otherwise, to get the conservative bound, one can set $p^{(+)j}_A=1$.
\item
$p^{(-)j}_A$, the probability that the event $j$ was initiated by the
primary $A$ with energy $E_0 \notin \{E\}$; in the case when the event
$j$ is unlikely to be induced by primary $A$, $p^{(-)j}_A$ is given by
Eq.~\eqref{estimate-bound} with $f_A({\bf c})= f^{(-)}_A({\bf c})$,
otherwise, to get the conservative bound, one can set $p^{(-)j}_A=0$.
\end{itemize}
Generally, these two probabilities do not sum up to unity,
$p^{(+)j}_A+p^{(-)j}_A<1$. To proceed further, one needs a conjecture
about other possibilities. Namely, we should distinguish between two other
probabilities,
\begin{itemize}
 \item
$p^{(+)j}_{\overline A}$, the probability that the event $j$ was initiated
by any other primary than $A$ with energy $E_0 \in \{E\}$;
\item
$p^{(-)j}_{\overline A}$, the probability that the event $j$ was initiated
by any other primary than $A$ with energy $E_0 \notin \{E\}$;
\end{itemize}
Without assumptions about possible primaries other than $A$, the two cases
$p^{(+)j}_{\overline A}$ and $p^{(-)j}_{\overline A}$ cannot be
distinguished by simulations. A reasonable solution is to assume loosely
that the energy was determined correctly by an experiment, $E_0=E_{\rm
rec}$, and the probability to have the primary energy $E_0$ follows the
distribution $g_E(E_{\rm obs},E_0)$. Then,
$$
p^{(+)j}_{\overline A}= \kappa _j \int \limits_{E \in \{E\}}\!
g_E(E^{}_{{\rm obs},j},E)\, dE,
~~~~~
p^{(-)j}_{\overline A}= \kappa _j \int \limits_{E \notin \{E\}}\!
g_E(E^{}_{{\rm obs},j},E)\, dE,
$$
where the normalisation factor $\kappa _j$ is determined from the
condition
$$
p^{(+)j}_A+p^{(-)j}_A+ p^{(+)j}_{\overline A}+p^{(-)j}_{\overline A} =1.
$$

Given the probabilities $p^{\pm(j)}_{A,\overline A}$ for each real event
$j$ in the sample, one can determine the probability ${\cal P}(n_1,n_2)$
to have, among $N$ observed events, $n_1$ initiated by primaries $A$ with
energies $E_0 \in \{E\}$ and $n_2$ initiated by any other primaries with
energies $E_0 \in \{E\}$. We are not interested in what happens outside
the domain $\{E\}$, so in what follows we do not distinguish $p^{(-)j}_A$
and  $p^{(-)j}_{\overline A}$ and use
$$
p^{(-)j}\equiv p^{(-)j}_{A} + p^{(-)j}_{\overline A} =
1-p^{(+)j}_{A}-p^{(+)j}_{\overline A}.
$$

The probability ${\cal P}(n_1,n_2)$ can be calculated as follows.
\footnote{Alternatively, a simple and fast
practical way to determine ${\cal P}(n_1,n_2)$ makes use of the
Monte-Carlo simulations. One generates a large number of sets of $N$
elements, each element marked either ``$A,\,E_0\in\{E\}$'', or
``${\overline A},\,E_0\in\{E\}$'', or ``$E_0\notin\{E\}$'' with the
probabilities $p_A^{(+)j}$, $p_{\overline A}^{(+)j}$ and $p^{(-)j}$,
respectively (these probabilities are different for different elements
$j=1,\dots N$). To get ${\cal P}(n_1,n_2)$, one simply counts the number
of sets with $n_1$ elements marked  ``$A,\,E_0\in\{E\}$'' and $n_2$
elements marked ``${\overline A},\,E_0\in\{E\}$'' and divides this number
by the total number of simulated sets.}
The probability to have $i_1$-th, \dots, $i_{n_1}$-th observed events
(to distinguish them, let us put them in order, $i_1<\dots< i_{n_1}$)
induced by primaries $A$ with $E\in\{E\}$
and $k_1$-th, \dots, $k_{n_2}$-th events ($k_1<\dots< k_{n_2}$, $i_j\neq
k_l$) induced by any other primaries with $E\in \{E\}$ is given
simply by the product
\[
{\cal P}\l \{i_j\}, \{k_l\}\r=
\prod_{i_j}p_A^{(+)i_j}\prod_{k_l}p_{\overline A}^{(+)k_l}
\prod_{m_n\ne i_j,k_l} p^{(-)m_n},
~~~~
1\le i_j,k_l,m_n \le N.
\]
To calculate ${\cal P}(n_1,n_2)$ one sums over
all possible ordered subsets $(\{i_j\}, \{k_l\})$
\begin{align}
{\cal   P}(n_1,n_2)&
=\sum_{{}^{i_1<i_2<\dots<i_{n_1}}_{k_1<k_2<\dots<k_{n_2}},i_j\neq
  k_l}{\cal P}\l \{i_j\}, \{k_l\}\r \nonumber \\
&=  \sum_{{}^{i_1<i_2<\dots<i_{n_1}}_{k_1<k_2<\dots<k_{n_2}},i_j\neq
  k_l} \left[\prod_{i_j}p_A^{(+)i_j}\prod_{k_l}p_{\overline A}^{(+)k_l}
  \prod_{m_n\ne i_j,\,k_l}
p^{(-)m_n}
\right]\;,
\label{P}
\end{align}
where $1\le i_j, k_l, m_n \le N$.

Let us suppose now that the fraction of primaries $A$ in the total flux of
particles with energies $E_0\in \{E\}$ is $\epsilon _A$.
By definition,
$\epsilon_A$ is the probability for a single shower with $E_0\in \{E\}$ to
be induced by $A$, while $(1-\epsilon_A)$ is a probability for a single
shower with $E_0\in \{ E\}$ to be induced by any other primary. Hence
${\cal P}(\epsilon)$, the probability that the observed results are
reproduced for a given $\epsilon _A$, can be expressed via ${\cal
P}(n_1,n_2)$ by making use of the formula for a conditional probability,
\begin{equation}
{\cal P}(\epsilon_A)=\sum_{n_1,n_2}^{n_1+n_2\leq N}
{\cal   P}(n_1,n_2)\epsilon_A^{n_1}\l1-\epsilon_A\r^{n_2}.
\label{8*}
\end{equation}
(cf.\ Ref.~\cite{Homola} for a particular case $n_1+n_2=N$).
The cases $n_1+n_2<N$ reflect the possibility that
$N-n_1-n_2$ events may correspond to primaries with $E_0\notin\{E\}$.

Making use of the function $P(\epsilon _A)$, one can constrain the
fraction $\epsilon _A$ at a given confidence level $\xi $. Indeed, the
allowed region of $\epsilon _A$ corresponds to
\begin{equation}
P(\epsilon _A)\ge 1-\xi.
\label{L*}
\end{equation}
For sufficiently small values of $p_A^{(+)j}$, the function
${\cal P}(\epsilon)$ is a monotonically
decreasing function and the allowed region corresponds to $\epsilon
_A<\epsilon _0$, where $\epsilon _0$ is determined by
${\cal P}(\epsilon_0)=1-\xi$;  we obtain an upper limit on
$\epsilon_A$.

Alternatively, $P(\epsilon _A)$ may be non-monotonic, and the
condition~(\ref{L*}) may determine one or several intervals allowed for
$\epsilon _A$ at the confidence level $\xi $. The most probable value of
$\epsilon _A$ corresponds to the maximal value of $P(\epsilon _A)$ over
$0\le \epsilon _A \le 1$.
 
Let us remind at this point that if the allowed interval for $\epsilon
_A$ does not include $\epsilon _A=0$ at a given C.L., this does not mean
we can be sure that there are $A$-primaries in Nature: this latter
statement is true only in the frameworks of the studied hypothesis
(primary $A$ versus any other primary with correct energy determination).

The limit one obtains in this way is however biased because some
particles with $E_0\in \{E\}$ may have reconstructed energies so
different from $E_0$ that the corresponding events would not enter the
experimental sample chosen for the study. To correct the final result
for these ``lost'' particles, another simulation is required. One
should simulate sufficiently large number $m$ of air showers initiated
by the primaries $A$ with arrival directions distributed randomly
according to the experimental acceptance and energies randomly chosen
from the domain $E_0\in \{E\}$, this time with the realistic spectrum
$E_0^{-\alpha}$.  Some
$m_{\rm lost}$ of these showers will have reconstructed energies
$E_{\rm rec}$ (or some other parameters) such that they would escape
from the sample chosen for the study; the fraction of these ``lost''
events is thus $\lambda =m_{\rm lost}/m$.
The true fraction
of $A$-primaries $\epsilon _{A,\,\rm true}$ is given by the ratio of
the number of $A$-induced events $M_{A,\,\rm true}$ and the sum
of the number of these events and the number of events initiated
by other particles $M$,
\begin{equation}
\label{add-add1}
\epsilon_{A,\,\rm true}=\frac{M_{A,\,\rm true}}{M_{A,\,\rm true}+M}\;.
\end{equation}
Similarly, the accessible to our study fraction of $A$-particles
$\epsilon_A $ is given by the ratio of accessible to our study
fraction of $A$-induced events $M_A=(1-\lambda) M_{A,\,\rm true}$ and
the sum of the number of these events and the number of events
initiated by other particles $M$,
\begin{equation}
\label{add-add2}
\epsilon_{A}=\frac{M_A}{M_A+M}=
\frac{(1-\lambda)M_{A,\,\rm true}}{(1-\lambda)M_{A,\,\rm true}+M}\;.
\end{equation}
Here we assume that other particles do not escape from the sample
chosen for the study. Equations \eqref{add-add1} and
\eqref{add-add2} enable one to place a bound on
\begin{equation}
\label{lost}
\epsilon _{A,\, \rm true} = \frac{\epsilon_A}{1-\lambda +\lambda \epsilon_A}
\end{equation}
 from previously obtained limit on $\epsilon _A$.

In the energy region to be studied,
the expected spectra of different primaries may be different.
Corresponding spectral indices may be
probed within our approach. Indeed, in the example considered above,
${\cal P}(\epsilon_A)$ depends also on the spectral index
$\alpha_A$, Eq.(\ref{w2}). The search for the
maximum of ${\cal P}(\epsilon_A)$ could be performed with account of
this new variable thus revealing, in general, the excluded (or most
probable, see sections below) region(s) of ($\epsilon_A$, $\alpha_A$)
parameter space. The extension to several primaries $A_1,A_2,\dots$ is
straightforward.

\subsubsection{Two or more primary types}
\label{sec:ensemble-2-primary}
Let us turn now to the study of the conjecture that all primary particles
in the energy domain $\{E\}$ were either $A_1$ or $A_2$. This requires
simulation and processing of the libraries of $A_1$- and $A_2$-induced
showers and no additional assumptions. The distributions
$f^{(\pm) j}_{A_1,A_2}({\bf c})$ are obtained for each event $j$ and four
probabilities are obtained:
\begin{itemize}
 \item
$p^{(+)j}_{A_1}$ --- event $j$ initiated by the primary $A_1$ with $E_0\in
\{E\}$;
\item
$p^{(-)j}_{A_1}$ --- event $j$ initiated by the primary $A_1$ with
$E_0\notin \{E\}$;
\item
$p^{(+)j}_{A_2}$ --- event $j$ initiated by the primary $A_2$ with $E_0\in
\{E\}$;
\item
$p^{(-)j}_{A_2}$ --- event $j$ initiated by the primary $A_2$ with
$E_0\notin \{E\}$
\end{itemize}
These probabilities are determined as follows,
\begin{equation}
p^{(\pm)j}_{A_1,A_2}={f^{(\pm) j}_{A_1,A_2}({\bf c}_{\rm obs}) \over
f^{(+) j}_{A_1} ({\bf c}_{\rm obs})+
f^{(-) j}_{A_1}({\bf c}_{\rm obs}) +
f^{(+) j}_{A_2}({\bf c}_{\rm obs}) +
f^{(-) j}_{A_2}({\bf c}_{\rm obs})}.
\label{N*}
\end{equation}
Now
\begin{equation}
p^{(-)j}=
p^{(-)j}_{A_1}+
p^{(-)j}_{A_2}=1-
p^{(+)j}_{A_1}-
p^{(+)j}_{A_2}
\label{N**}
\end{equation}
corresponds to the primaries with $E_0\notin \{E\}$.

The probability ${\cal P}(n_1,n_2)$ corresponds now to the case when in
the set, $n_1$ primaries $A_1$ with $E_0\in E$, $n_2$ primaries $A_2$ with
$E_0\in E$ and $N-n_1-n_2$ primaries $A_1$ or $A_2$ with $E_0\notin \{E\}$
are present. It is determined by the same Eq.~(\ref{P}) with the
replacements $A\to A_1$, ${\overline A}\to A_2$, where
$p^{(+)j}_{A_{1,2}}$ and $p^{(-)j}$ are now given by Eqs.~(\ref{N*}),
(\ref{N**}). It also can be determined by the Monte-Carlo simulation
described above.

If the fraction of primaries $A_1$ in the energy domain $\{E\}$ is
$\epsilon _{A_1}$, then, within the conjecture we study, the rest
$1-\epsilon _{A_1}$ correspond to $A_2$.  Thus $P(\epsilon _{A_1})$ is
given by the same Eq.~(\ref{8*}) and the region of $\epsilon _{A_1}$
allowed at the confidence level $\xi $ is again determined by $P(\epsilon
_{A_1})\ge 1-\xi $.

The most probable value of $\epsilon _{A_1}$ may be determined by
maximization of $P(\epsilon _{A_1})$.

Account of the ``lost'' events requires, in this case, determination of
two ``lost'' fractions, $\lambda _{A_1}$ and $\lambda _{A_2}$. From the
relation
$$
\epsilon _{A_1}= {\epsilon _{A_1}^{\rm true} (1-\lambda _{A_1})\over
\epsilon _{A_1}^{\rm true} (1-\lambda _{A_1}) +
(1-\epsilon _{A_1}^{\rm true}) (1-\lambda _{A_2})}
$$
one obtains
\begin{equation}
\epsilon _{A_1}^{\rm true} = {\epsilon _{A_1} (1-\lambda _{A_2}) \over 1 -
\lambda _{A_1}+\epsilon _{A_1}(\lambda _{A_1}-\lambda _{A_2})}.
\label{lost:2-primaries}
\end{equation}

Generalization of the procedure and equations to the case of more than
two primary
types $A_k$, $k=1,\dots K$, $K>2$, is straightforward.
In particular, one defines the probability to
have a set of fractions $(\epsilon_{A_1},\dots,\epsilon_{A_K})$,
$\sum_k\epsilon_{A_k}=1$, among the primaries with $E_0\in\{E\}$.
The probability ${\cal P}(n_1,n_2)$ is now replaced by ${\cal
P}(n_1,\dots,n_K)$ which can be calculated using a similar method.
The function
$P(\epsilon_{A_1},\dots,\epsilon_{A_K})$ can be again expressed via ${\cal
P}(n_1,\dots,n_K)$ by making use of the formulae for conditional
probability,
\[
P(\epsilon_{A_1},\dots,\epsilon_{A_K})
=\sum_{{}^{n_1,\dots,n_K}_{\l\sum_i n_i\r\leq N}}
{\cal   P}(n_1,\dots,n_K)\cdot\prod_{k=1}^{K} \epsilon_{A_i}^{n_k}\;,~~~~
\sum_{k=1}^{K}\epsilon_{A_k}=1\;.
\]
Determination of the best-fit
composition would require, in this case, maximization of $P(\epsilon
_{A_1}, \dots, \epsilon _{A_K-1})$ with respect to $K-1$ fractions
$\epsilon _k$ (note that $\epsilon _{A_K}=1-\sum_{k=1}^{K-1} \epsilon
_{A_k}$).

\section{Examples}
\label{sec:example}
In this section, we illustrate the method with several simple (toy) examples.
For the data, we choose the highest-energy cosmic-ray events reported by
the AGASA~\cite{AGASA:events} and Yakutsk~\cite{Yakutsk:events}
experiments (see Table~\ref{tab:events} for the list of events we use
and Ref.~\cite{gamma-limit} for more details). The E-observable used by
the
both experiments is $S(600)$, the signal density at 600~m from the shower
core. For a given zenith angle $\theta$, it is in one-to-one
correspondence with $E_{\rm rec}$. Here, we use only one c-observable,
$\rho _\mu (1000)$, the muon density at 1000~m from the core. Observed
arrival directions, $E_{\rm obs}$ and $\rho _\mu (1000)$ are given in
Table~\ref{tab:events} in Appendix~\ref{app:events}, where we also present
technical details of the simulation procedure. The procedure used to
reconstruct $E_{\rm rec}$ and $\rho _\mu (1000)$ is described in
Appendix~\ref{app:reconstruction}.

We do not discuss many experimental details and
the model dependence of the results, so all three examples in this section
should be considered as toy ones serving the only purpose to illustrate
the method.

\subsection{Example 1: an individual event}
\label{sec:example-event}
The procedure described in this section has to be performed for each real
event in the dataset. Shower libraries for different primaries are
generated as described in Appendix~\ref{app:events}, and $E_{\rm rec}$ and
$\rho _\mu (1000)$  are determined for each simulated shower using the
procedure presented in Appendix~\ref{app:reconstruction}. In this example,
we consider the highest-energy AGASA event (event 1 from
Table~\ref{tab:events}).

Simulations of the AGASA detector and the reconstruction
procedure~\cite{AGASA:events} indicate that the function $g_E$ is Gaussian
in $\log E$ with the width corresponding to the standard deviation of 25\%
of the central value in terms of energy. This corresponds to $\sigma_l
\approx 0.104$ for $\log E$,
\[
g_E(E_{\rm obs},E_{\rm rec})=\frac{1}{\sqrt{2\pi}\sigma_l}
\e^{-\frac{\log^2(E_{\rm rec}/E_{\rm obs})}{2\sigma_l^2}}\;,~~~~~
\int_{0}^{+\infty}g_E(E_{\rm obs},E_{\rm rec})\frac{dE_{\rm
rec}}{E_{\rm rec}}=1\;.
\]

\begin{figure}
\centerline{
\includegraphics[width=112mm]{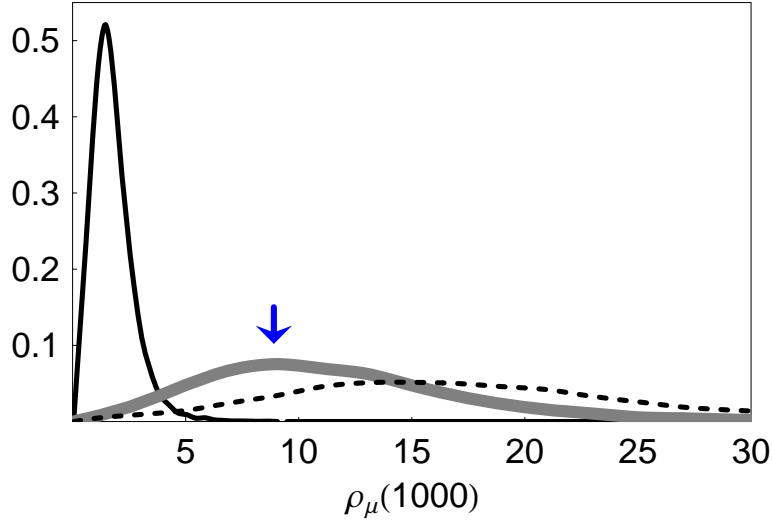}
}
\caption{
Distributions of muon densities $f_A$ of simulated events consistent with
Event~1 by the arrival direction and the reconstructed energy: thin dark
line, $A=\gamma $; thick gray line, $A=p$; dashed line, $A=$Fe. The arrow
indicates the observed value of $\rho _\mu (1000)$ for Event~1.
\label{Fig:f_A}
}
\end{figure}

Though the detector errors in determination of $\rho _\mu $ and fitting
the muon LDF may vary with the muon density and number of detectors hit,
in our toy example we use for the function $g_c$ the
approximation~\cite{AGASAmu} of the Gaussian distribution with the width
of 40\% of the central value. In our case, there is only one C-observable,
$\rho _\mu (1000)\equiv c$, and we thus have
\[
g_c(c',c)=\frac{\e^{-\frac{\l c-c'\r^2}{2 \sigma _c^2}}}
{\int_0^\infty \e^{-\frac{\l c-c'\r^2}{2 \sigma _c^2}} dc },
~~~~
\sigma _c=0.4c'.
\]
If we are interested in the primary particle of this particular event
only, we should use Eq.~(\ref{f_A(c)}) to determine $f_\gamma $, $f_p$ and
$f_{\rm Fe}$. These functions are plotted in Fig.~\ref{Fig:f_A}.
These three functions allow to quantify, in particular, the answers to
the questions:
\begin{itemize}
 \item
How bad this event is described by a photon (of any energy), without
fixing other possibilities? With the help of Eqs.~(\ref{p_A-tilde}),
(\ref{estimate-bound}), one finds for this case
$$
\tilde p_\gamma \approx p_\gamma \approx 0.
$$
\item
Suppose that the event may be initiated either by a photon or by a proton
(of any energy). Which primary is prefered? Application of
Eq.~(\ref{p_A1,A2}) results in
$$
p_\gamma \approx 0.001; ~~~~ p_p\approx 0.999.
$$
\item
Suppose that the event may be initiated either by a proton or by an iron
nuclei (of any energy). Which primary is prefered? Application of
Eq.~(\ref{p_A1,A2}) results in
$$
p_p \approx 0.69; ~~~~ p_{\rm Fe}\approx 0.31.
$$
\end{itemize}
 
We emphasise once again that the probability $p_p\approx 1$ in the
$p$-$\gamma $ comparison does not mean it was surely a proton: this is
true only for the hypothesis that either protons or photons can be the
primaries. Indeed, $p_p<1$ for the $p$-Fe comparison. The difference
between protons and iron nuclei is not significant for a single
event(cf.~Fig.~\ref{w2}). Nevertheless, this difference becomes more
pronounced for samples of several events.

We note that the two-primary comparison is complementary to the
one-primary analyses because the latters assume correct energy
determination for a part of showers (this or any other hypothesis is
necessary for obtaining quantitative results).  

\begin{figure}
\centerline{
\includegraphics[width=112mm]{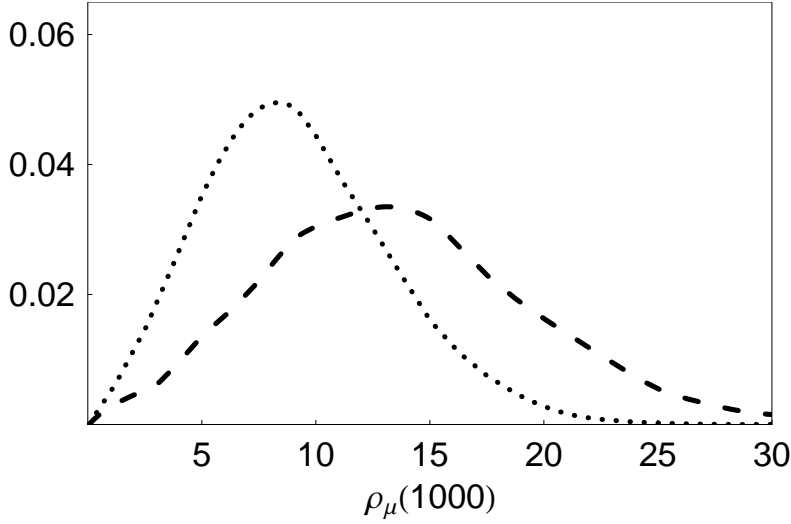}
}
\caption{
Distributions of muon densities $f_p$ of simulated proton-induced events
consistent with Event~1 by the arrival direction and the reconstructed
energy: dotted line, $f_p^{(-)}$ (corresponds to $E_0<1.5
\cdot 10^{20}$~eV); dashed line, $f_p^{(+)}$ (corresponds to
$E_0>1.5 \cdot 10^{20}$~eV).
\label{Fig:f_p+-}
}
\end{figure}

In the following examples (Sec.~\ref{sec:example-gamma-limit},
\ref{sec:example-p-Fe}), we will study ensembles of events containing this
Event~1. Hence we will need the functions $f_A^{(\pm)}$ determined
in Eq.~(\ref{f+-}) for the energy domains $\{E\}$ to be used in these
examples: $E_0\ge 10^{20}$~eV for Example~2
(Sec.~\ref{sec:example-gamma-limit}) and $E_0\ge 1.5\cdot 10^{20}$~eV for
Example~3 (Sec.~\ref{sec:example-p-Fe}). For the latter case, we plot in
Fig.~\ref{Fig:f_p+-} functions $f_p^{(\pm)}$; the function $f_p^{(-)}$
corresponds to lower energies and hence lower muon densities.
The probabilities $p_\gamma ^{(\pm)}$ and $p_{\overline \gamma} ^{(\pm)}$ for
the limit on the gamma-ray primaries, $\{E\}=\{E_0>10^{20}~{\rm eV}\}$,
calculated as discussed in Sec.~\ref{sec:ensemble-1-primary}, and
probabilities $p_p^{(\pm)}$ and $p_{\rm Fe}^{(\pm)}$, $\{E\}=\{E_0>1.5 \cdot
10^{20}~{\rm eV}\}$, calculated with the help of Eq.~(\ref{N*}) for the
proton-iron discrimination, are presented both for the Event~1 discussed
here and for other events in the samples in Tables~\ref{tab:p_gamma} and
\ref{tab:p_p-Fe} in Sec.~\ref{sec:example-gamma-limit} and
\ref{sec:example-p-Fe}, respectively.

\subsection{Example 2: Limit on the gamma-ray fraction}
\label{sec:example-gamma-limit}
We  concentrate here on obtaining the limit on the gamma-ray primaries
with energies $E_0>10^{20}$~eV based on the AGASA data. The energy domain
of interest is thus
$\{E\}=\{E_0>10^{20}~{\rm eV}\}$.
For the experimental sample we take all AGASA events with known muon data
and energies $E_{\rm obs}>8 \cdot 10^{19}$~eV: this widening of the energy
range (cf.\ Sec.~\ref{sec:general-ensemble}) compared to the domain
$\{E\}$ is justified because photon energies may be estimated incorrectly
by the experiment. There are six events in the sample (events 1--6 in
Table~\ref{tab:events}); see Ref.~\cite{gamma-limit} for the application
of the same procedure to a larger dataset supplemented by Yakutsk events. For
each of the real events we determine the probabilities $p_{\gamma ,{\overline
\gamma }}^{(\pm)}$ in the same way as described in
Sec.~\ref{sec:example-event} for the Event~1. The probabilities are given
in Table~\ref{tab:p_gamma}.
\begin{table}
\begin{tabular}{ccccc}
\hline
Event & $p_\gamma ^{(+)}$ & $p_\gamma ^{(-)}$ & $p_{\overline\gamma} ^{(+)}$
& $p_{\overline\gamma} ^{(-)}$\\
\hline
1 & 0.000 & 0.000 & 1.000 & 0.000\\
2 & 0.001 & 0.000 & 0.998 & 0.001\\
3 & 0.013 & 0.003 & 0.921 & 0.063\\
4 & 0.003 & 0.000 & 0.887 & 0.111\\
5 & 0.000 & 0.000 & 0.580 & 0.420\\
6 & 0.000 & 0.000 & 0.565 & 0.435\\
\hline
\end{tabular}
\vspace{5pt}
\caption{
Probabilities $p_{\gamma ,{\overline \gamma }}^{(\pm)}$ (see text for
notations) for individual events 1--6 entering the sample for Example 2
(the limit on the gamma-ray primary fraction with
$\{E\}=\{E_0>10^{20}~{\rm eV}\}$
).
\label{tab:p_gamma}}
\end{table}
Corresponding probabilities ${\cal P}(n_1,n_2)$ to have $n_1$ gamma-ray
primaries with $E_0>10^{20}$~eV and $n_2$ other primaries with
$E_0>10^{20}$~eV, calculated as discussed in
Sec.~\ref{sec:ensemble-1-primary}, are
presented in Table~\ref{tab:P(i)(j)_gamma}.
\begin{table}
\begin{tabular}{cccccccc}
\hline
& $n_2=0$ & $n_2=1$ & $n_2=2$ & $n_2=3$ & $n_2=4$ & $n_2=5$ &
$n_2=6$\\
\hline
$n_1$=0 & 0.000 & 0.000 & 0.001 & 0.036 & 0.234 & 0.442 & 0.268\\
$n_1$=1 & 0.000 & 0.000 & 0.000 & 0.005 & 0.009 & 0.005 &      \\
$n_1$=2 & 0.000 & 0.000 & 0.000 & 0.000 & 0.000 &       &      \\
$n_1$=3 & 0.000 & 0.000 & 0.000 & 0.000 &       &       &      \\
$n_1$=4 & 0.000 & 0.000 & 0.000 &       &       &       &      \\
$n_1$=5 & 0.000 & 0.000 &       &       &       &       &      \\
$n_1$=6 & 0.000 &       &       &       &       &       &      \\
\hline
\end{tabular}
\vspace{5pt}
\caption{
The probabilities ${\cal P}(n_1,n_2)$ to have $n_1$ gamma-ray primaries
with $E_0>10^{20}$~eV and $n_2$ other primaries with $E_0>10^{20}$~eV in
the sample of 6 events used in Example 2. Correct experimental energy
determination was assumed for non-photon primaries. Note that $n_1+n_2\le
6$ and the rest $6-n_1-n_2$ particles have energies $E_0<10^{20}$~eV in
each case.
\label{tab:P(i)(j)_gamma}
}
\end{table}

\begin{figure}
\centerline{
\includegraphics[width=112mm]{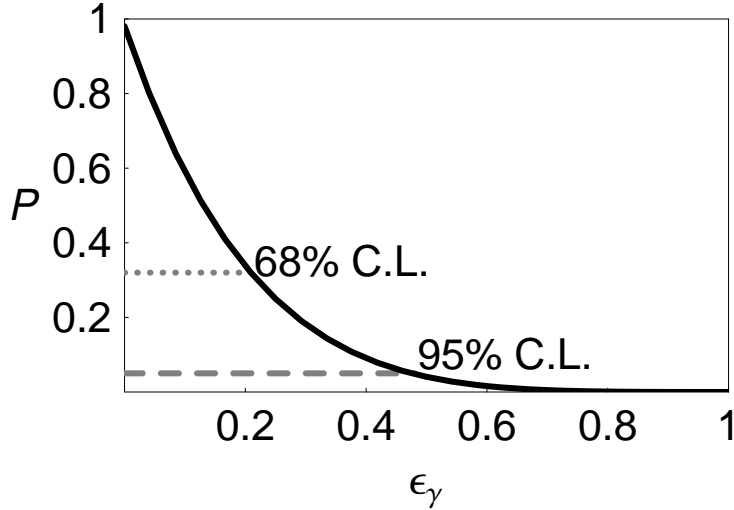}
}
\caption{
$P(\epsilon _\gamma )$ calculated for the Example~2. Dashed line
corresponds to $1-P=0.95$ (the 95\%~C.L. upper limit on $\epsilon _\gamma
$), dotted line -- to $1-P=0.68$ (the 68\%~C.L. upper limit).
\label{Fig:P(epsilon_gamma)}
}
\end{figure}

Figure~\ref{Fig:P(epsilon_gamma)} presents the function $P(\epsilon
_\gamma )$ obtained from the values of ${\cal P}(n_1,n_2)$ with the help
of Eq.~(\ref{8*}).
It is monotonically decreasing: this reflects the fact that the presence
of gamma-ray primaries is disfavoured and the most probable fraction
(corresponding to the maximal value of $P(\epsilon _\gamma )$)
is $\epsilon _\gamma =0$. We illustrate Eq.~(\ref{L*}) for
two confidence levels, $\xi =0.95$ and $\xi=0.68$: $ \epsilon _\gamma <
0.21 $ at 68\%~C.L.\ and $ \epsilon _\gamma < 0.47 $ at 95\%~C.L.

To account for the ``lost photons'' whose true energies were
$E_0>10^{20}$~eV but reconstructed energies were $E_{\rm rec}<8\cdot
10^{19}$~eV, we simulated $m=1000$ photon-induced showers whose arrival
directions were distributed following the AGASA acceptance (with the
zenith angle cut of 45$^\circ$) and energies $E_0\ge 10^{20}$~eV chosen
randomly to follow $E_0^{-2}$ spectrum suggested by several theoretical
models with raising super-GZK photon component. The fraction $\lambda
\approx 0.035$ of these events had reconstructed energies $E_{\rm
rec}<8\cdot 10^{19}$~eV and are therefore ``lost''. Application of
Eq.~(\ref{lost}) results in
$$
\epsilon _{\gamma, {\rm true}} < 0.22 ~~~
{\rm at~68\%~C.L.},
$$
$$
\epsilon _{\gamma, {\rm true}} < 0.50 ~~~
{\rm at~95\%~C.L.}
$$

\subsection{Example 3: Favoured hadronic composition}
\label{sec:example-p-Fe}
To illustrate the case when two possible primary types are compared, we
consider four events with $E_{\rm obs}>1.5\cdot 10^{20}$~eV and known
muon content (two observed by AGASA and two by Yakutsk; events 1, 2, 7
and 8 from Table~\ref{tab:events}). The primary composition in the
domain $\{E\}=\{E_0>1.5 \cdot 10^{20}~{\rm eV}\}$ will be constrained
within the hypothesis that the primaries are either protons or iron
nuclei.

\begin{figure}
\centerline{
\includegraphics[width=112mm]{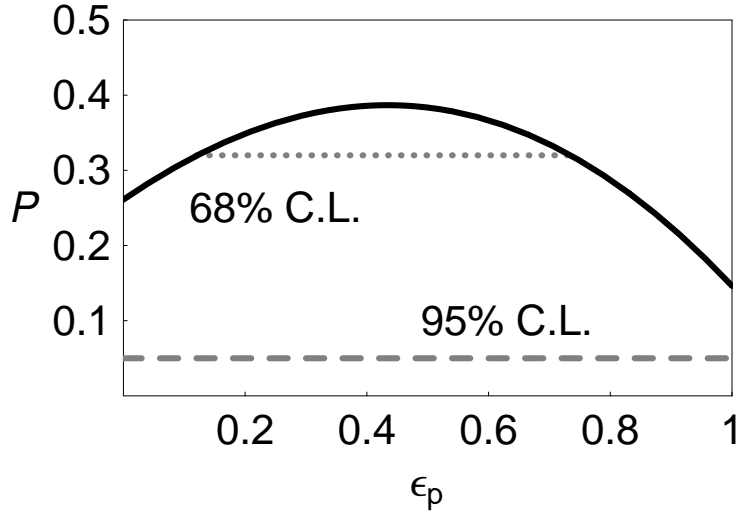}
}
\caption{
$P(\epsilon _p )$ calculated for the Example~3. Notations are the same as
in Fig.~\ref{Fig:P(epsilon_gamma)}
\label{Fig:P(epsilon_p)}
}
\end{figure}

In a way similar to Sec.~\ref{sec:example-event}, we process all four
events and obtain probabilities $p_{p, {\rm Fe}}^{(\pm)}$, Eq.~(\ref{N*}),
which are listed in Table~\ref{tab:p_p-Fe}.
\begin{table}
\begin{tabular}{ccccc}
\hline
Event & $p_p ^{(+)}$ & $p_p ^{(-)}$ & $p_{\rm Fe} ^{(+)}$
& $p_{\rm Fe} ^{(-)}$\\
\hline
1 & 0.254 & 0.445 & 0.136 & 0.165 \\
2 & 0.295 & 0.349 & 0.135 & 0.221 \\
7 & 0.163 & 0.001 & 0.735 & 0.101 \\
8 & 0.407 & 0.107 & 0.256 & 0.230 \\
\hline
\end{tabular}
\vspace{5pt}
\caption{
Probabilities $p_{p, {\rm Fe}}^{(\pm)}$ (see text for
notations) for individual events 1, 2, 7, 8 entering the sample for Example
3 (comparison of protons and iron nuclei).
\label{tab:p_p-Fe}}
\end{table}
They are transformed into probabilities
${\cal P}(n_1,n_2)$ to have $n_1$ proton primaries
with $E_0>1.5 \cdot 10^{20}$~eV and $n_2$ iron primaries with
$E_0>1.5 \cdot 10^{20}$~eV
among the four events we discuss (see Table~\ref{tab:P(i)(j)_p-Fe}).
\begin{table}
\begin{tabular}{cccccc}
\hline
 & $n_2=0$ & $n_2=1$ & $n_2=2$ & $n_2=3$ & $n_2=4$ \\
\hline
$n_1$=0 & 0.012 & 0.102 & 0.110 & 0.035 & 0.003 \\
$n_1$=1 & 0.044 & 0.226 & 0.138 & 0.020 &       \\
$n_1$=2 & 0.057 & 0.149 & 0.041 &       &       \\
$n_1$=3 & 0.028 & 0.030 &       &       &       \\
$n_1$=4 & 0.005 &       &       &       &       \\
\hline
\end{tabular}
\vspace{5pt}
\caption{
The probabilities ${\cal P}(n_1,n_2)$ to have $n_1$ proton primaries
with $E_0>1.5 \cdot 10^{20}$~eV and $n_2$ iron primaries with
$E_0>1.5 \cdot 10^{20}$~eV in the sample of 4 events used in Example 3.
Note that $n_1+n_2\le 4$ and $4-n_1-n_2$ particles
have energies $E_0<1.5\cdot 10^{20}$~eV in each case.
\label{tab:P(i)(j)_p-Fe}}
\end{table}
The function $P(\epsilon _p)$, plotted in Fig.~\ref{Fig:P(epsilon_p)},
looks quite different from $P(\epsilon _\gamma )$ from the previous
example. There is no clear preference of either proton or iron
primaries, and the function has a wide bump. The maximum corresponds to
the most probable proton fraction $\epsilon _p\sim 0.39$, but at the
95\%~C.L., the proton fraction is unconstrained.
Clearly, the main reason for the weakness of the constraint is
the small number of events in the sample. The limit at the 68\%~C.L.
illustrates the situation expected at higher confidence level for larger
samples: the constraint reads $0.12<\epsilon _p<0.73$. It is reassuring that
with only four events and for such a difficult task as to distinguish
between different hadronic primaries, our method still allows to obtain
some indicative information: heavier composition is slightly prefered.

A minor additional complication in this example, as compared to
the Example~2, relates to the calculation of the fractions of ``lost''
particles. The reason is that we unified data from two experiments in a
single sample. We suppose that both experiments have similar
sensitivities to different primaries (otherwise the unification of data
sets would not be justified anyway). Then the {\rm expected} total number
of events which may be observed by an experiment is proportional
to its exposure. We denote by $\epsilon _p^{{\rm true}(I)}$ and
$\epsilon _p^{{\rm true}(II)}$ the proton fraction calculated with the
help of Eq.~(\ref{lost:2-primaries}) for the experiments I and II whose
exposures are $a_I$ and $a_{II}$, respectively. The fraction $\epsilon _p$
estimated for the combined dataset is then
$$
\epsilon _p^{\rm true}=
{\epsilon _p^{{\rm true}(I)}\over 1+a_{(II)}/a_I}+
{\epsilon _p^{{\rm true}(II)}\over 1+a_I/a_{(II)}}.
$$
From the simulation of 1000 proton-induced and 1000 iron-induced showers
from various arrival directions for each of the two cases
-- (I) AGASA, $\theta<45^\circ$, and (II) Yakutsk, $\theta<60^\circ$ --
one obtains $\lambda _p^{(I)}\approx 0.05$, $\lambda _{\rm
Fe}^{(I)}\approx 0.06$, $\lambda _p^{(II)}\approx 0.11$, $\lambda _{\rm
Fe}^{(II)}\approx 0.08$. Since $\lambda _p$ and $\lambda _{\rm Fe}$ are
relatively close to each other, their contributions balance each other in
Eq.~(\ref{lost:2-primaries}), and the resulting limits on $\epsilon
_p^{\rm true}$ do not differ from the limits on $\epsilon _p$ without the
correction for ``lost'' events (clearly, this would not be so in a general
case).

\subsection{Testing the method with artificial samples}
\label{sec:example-artificial}
To test how well the method works, we perform analyses similar to those of
Sec.~\ref{sec:example-event}--\ref{sec:example-p-Fe} but for a simulated
data sample with known primaries. To this end, we generated showers with
the arrival directions of events 1--8 (this allowed us to use the shower
libraries generated for examples 1--3 above) and kept record of their
primary particles and $E_0$. For each shower, we reconstructed $S(600)$
(and hence $E_{\rm rec}$) and $\rho _\mu (1000)$ taking into account
random detector errors. Then, we performed our analyses and compared their
results with known primary content of the fake sample.

The events of the simulated sample are listed in
Table~\ref{tab:fake-events}. For the study similar to
Sec.~\ref{sec:example-gamma-limit}, we take events F1--F6 of which three
are initiated by protons and three by photons (that is, $\epsilon _\gamma
=0.5$). Individual probabilities $p_\gamma ,p_{\bar\gamma}$ for these
events are given in Table~\ref{tab:p_gamma-artificial}.
\begin{table}
  \begin{tabular}{ccccc}
\hline
Event & $p_\gamma ^{(+)}$ & $p_\gamma ^{(-)}$ & $p_{\overline\gamma} ^{(+)}$
& $p_{\overline\gamma} ^{(-)}$\\
\hline
F1 & 0.000 & 0.000 & 1.000 & 0.000\\
F2 & 0.000 & 0.000 & 0.993 & 0.007\\
F3 & 0.391 & 0.287 & 0.130 & 0.192\\
F4 & 0.000 & 0.000 & 0.887 & 0.113\\
F5 & 0.150 & 0.534 & 0.258 & 0.058\\
F6 & 0.273 & 0.090 & 0.630 & 0.007\\
\hline
\end{tabular}
\vspace{5pt}
\caption{
Probabilities $p_{\gamma ,{\overline \gamma }}^{(\pm)}$ (see text for
notations) for individual events F1--F6 entering the artificial sample for
the limit on the gamma-ray primary fraction with
$\{E\}=\{E_0>10^{20}~{\rm eV}\}$.
\label{tab:p_gamma-artificial}}
\end{table}
Application of the method gives, at the 95\%~C.L., $\epsilon _\gamma
<0.73$ if we take the sample with $E_{\rm obs}\ge 0.9 \cdot 10^{20}$~eV and
$\epsilon _\gamma <0.72$ if we use only five events with $E_{\rm obs}\ge
10^{20}$~eV (the difference in the fractions of lost photons
compensates the difference in statistics).

For a study similar to Example~2, we take four events (F1, F2, F7 and F8)
of which two are initiated by protons and two by iron nuclei, that is
$\epsilon _p=0.5$. The probabilities $P_{p,{\rm Fe}}^{(\pm)}$ are given in
Table~\ref{tab:p_p-Fe-artificial}.
\begin{table}
  \begin{tabular}{ccccc}
\hline
Event & $p_p ^{(+)}$ & $p_p ^{(-)}$ & $p_{\rm Fe} ^{(+)}$
& $p_{\rm Fe} ^{(-)}$\\
\hline
F1 & 0.281 & 0.084 & 0.450 & 0.184 \\
F2 & 0.159 & 0.421 & 0.086 & 0.333 \\
F7 & 0.159 & 0.064 & 0.300 & 0.476 \\
F8 & 0.458 & 0.070 & 0.323 & 0.149 \\
\hline
\end{tabular}
\vspace{5pt}
\caption{
Probabilities $p_{p, {\rm Fe}}^{(\pm)}$ (see text for
notations) for individual events F1, F2, F7, F8 entering the
artificial sample for comparison of protons and iron nuclei.
\label{tab:p_p-Fe-artificial}}
\end{table}
The most favoured composition is obtained to be $\epsilon _p\approx 0.45$,
but any is allowed at the 95\%~C.L.

\section{Conclusions}
\label{sec:concl}
To summarise, we presented a simple method which allows to improve
precision of the studies of the primary composition of ultra-high-energy
cosmic rays. Each event is studied individually. The simulated showers
are selected by the physical
observables to be consistent with the real event.

\begin{figure}
\centerline{
\includegraphics[width=112mm]{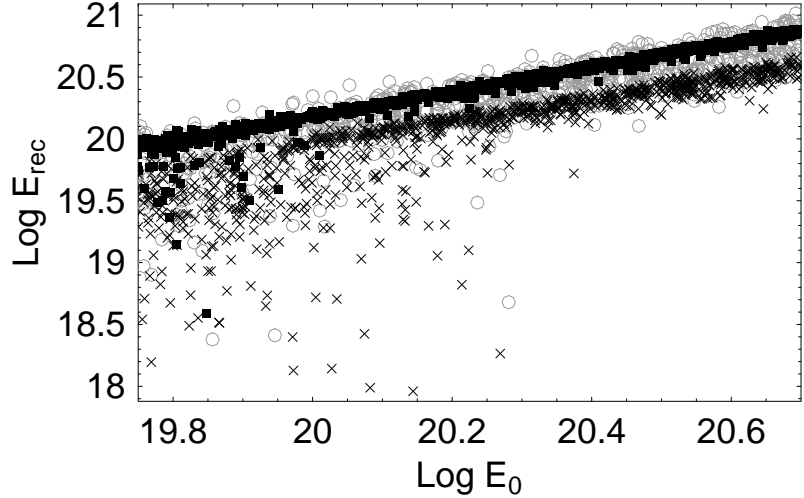}
}
\caption{
Direction dependence of the reconstructed energy for gamma-ray primaries.
Plotted is the reconstructed energy (determined by the AGASA method from
$S(600)$) versus the primary energy. Dark boxes: arrival direction of
the event 1; crosses: arrival direction of the event 3; grey circles:
arrival directions randomly distributed according to the AGASA acceptance
($0<\theta<45^\circ$).
Both
$E_0$ and $E_{\rm rec}$ are measured in eV.
\label{Fig:discussion-gamma}
}
\end{figure}

\begin{figure}
\centerline{
\includegraphics[width=112mm]{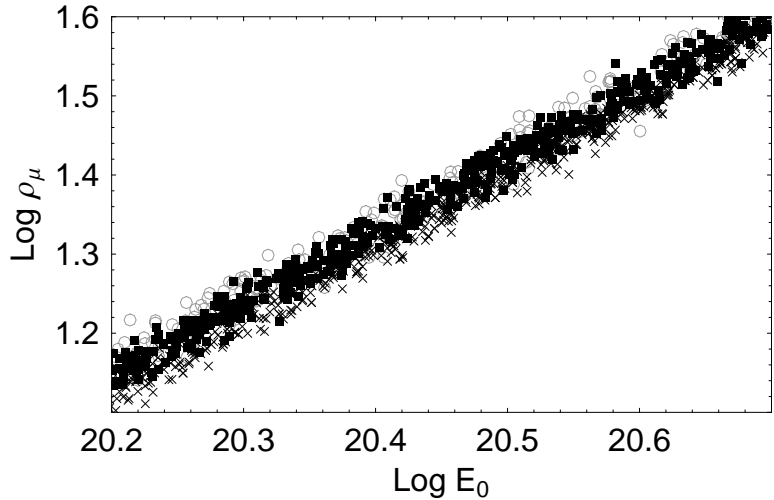}
}
\caption{
Direction dependence of muon density for iron primaries.
Plotted is $\rho _\mu (1000)$
versus the primary energy. Dark boxes: arrival direction of
the event 7 ($\theta \approx 48^\circ$); crosses: arrival direction of the
event 8 ($\theta\approx 59^\circ$); grey circles: arrival directions
randomly distributed according to the Yakutsk acceptance
($0<\theta<60^\circ$).
$E_0$ is measured in eV, $\rho _\mu $ is measured in m$^{-2}$.
\label{Fig:discussion-Fe}
}
\end{figure}

Our approach exploits the fact that
the uncertainty in discrimination of the primary-particle types by
conventional methods is determined not only by intrinsic fluctuations of
the shower development, but also by the spread related to the arrival
direction.
Averaging over arrival directions introduces artificial, easy-to-avoid,
fluctuations in the determination of both the reconstructed energy (see
Fig.~\ref{Fig:discussion-gamma} for an illustration) and
composition-related parameters (Fig.~\ref{Fig:discussion-Fe}).

While the direction-related systematical effects are controlled within
our method, there are still many other sources of uncertainty,
the most important among them are the low precision of the measurement of
composition-related quantities in experiments and artificial fluctuations
in simulated air showers due to use of the thinning approximation.
Use of multiple c-observables, their careful choice and optimisation of
simulations may help to improve further the precision of our method.

As for any other method based on air-shower simulations, the capabilities
of our approach are limited (and severely limited in the case of
distinguishing different hadronic primaries) because of persistent
theoretical problems in modelling high-energy hadronic interactions.  

As compared to the traditional methods, the event-by-event study requires
more computational time, which makes it difficult to implement it for
large samples in the form described here. However, almost the same level
of precision may be reached with reasonable binning in zenith (and in some
cases also azimuth) angle.

As we have seen in Sec.~\ref{sec:example}, the method is really strong in
studies of the small samples, notably for the highest-energy or
somewhat special (e.g.\ correlated with particular hypothetical sources)
events, where statistics is insufficient to obtain confident conclusions
with the help of traditional methods. In some cases, e.g.\ to constrain
the gamma-ray primaries, even a single event may be succesfully studied.

\ack We are indebted to L.~Dedenko, M.~Kachelrie\ss, M.~Pravdin,
K.~Shinozaki, D.~Semikoz and M.~Teshima for numerous helpful
discussions.
This work was supported in part by the INTAS grant 03-51-5112, by
the Russian Foundation of Basic Research grants 05-02-17363 (DG and GR)
and 04-02-17448 (DG), by the
grants of the President of the Russian Federation NS-7293.2006.2
(government contract 02.445.11.7370; DG, GR and ST) and MK-2974.2006.2
(DG), by the fellowships of the "Dynasty" foundation (awarded by the
Scientific Council of ICFPM, DG and GR) and of the Russian Science Support
Foundation (ST). Numerical part of the work was performed at the computer
cluster of the Theoretical Division of INR RAS.

\appendix
\section{Notations}
\label{app:notations}
$A$: the index enumerating different primaries, $A=p,\ {\rm Fe},\ \gamma
, \dots$

$\overline A$: denotes any other primary but $A$.

${\bf c}=(c_1,c_2,\dots)$ -- the set of c-observables (observables used
to distinguish different primaries). ${\bf c}_{\rm obs}$ denotes the
values of these observables for a particular real event.

E-observables -- the observables used for the energy estimation.

$E_0$ -- the real energy of the primary particle; for simulated events it
is the parameter of the simulation.

$E_{\rm rec}$ -- the reconstructed energy of a simulated shower;
it is an E-observable (in the sense that
it is in one-to-one correspondence with
the observable used to estimate the energy for a given arrival
direction) and can be determined for each air shower
according to the reconstruction procedure fixed by the
experiment (see Appendix~\ref{app:reconstruction}). In general, $E_{\rm
rec} \ne E_0$  due to both statistical and systematic errors.

$E_{\rm obs}$ -- the reported energy of a given observed air shower.

$\{ E \}$ -- the region of primary energies $E_0$ for which the
composition is to be constrained.

$\epsilon _A$ -- the fraction of the primaries $A$ in the total
flux of cosmic rays with energies $E_0\in \{ E \}$.

$\epsilon _{A,{\rm true}}$ -- the fraction $\epsilon _A$ corrected for the
``lost events''.

$\lambda $ -- the fraction of the ``lost events'', that is of the showers
whose primaries had the energies $E_0\in \{ E \}$ but which did not enter
the sample because of incorrect energy determination.

\section{Observed and simulated events used in examples}
\label{app:events}
Experimental information required for the simulations in Examples 1--3 of
Sec.~\ref{sec:example} is taken from
Refs.~\cite{gamma-limit,AGASA:events,Yakutsk:events} and is summarised in
Table~\ref{tab:events}.
\begin{table}
\begin{tabular}{cccccccc}
\hline
No.& Exp.&$E_{\rm obs}$&$\theta$&$\phi$& $\rho^{\rm obs} _\mu (1000)$ &
\multicolumn{1}{c}{Examples} & \multicolumn{1}{c}{Simulated} \\
(1)&(2)&(3)&(4)&(5)&(6)&(7)&(8)\\
\hline
1 &AGASA  & 2.46 & 36.5& 79.2 & 8.9 &1,2,3 &$\gamma $, $p$, Fe \\
2 &AGASA  & 2.13 & 22.9& 55.5 &10.7 &2,3   &$\gamma $, $p$, Fe  \\
3 &AGASA  & 1.44 & 14.2& 27.5 & 8.7 &2     &$\gamma $ \\
4 &AGASA  & 1.34 & 35.1&234.9 & 5.9 &2     &$\gamma $ \\
5 &AGASA  & 1.05 & 33.7&291.6 &12.6 &2     &$\gamma $ \\
6 &AGASA  & 1.04 & 35.6&100.0 & 9.3 &2     &$\gamma $ \\
7 &Yakutsk& 1.60 & 47.7&180.8 &19.6 &2,3   &$p$, Fe  \\
8 &Yakutsk& 1.50 & 58.7&230.6 &11.8 &2,3   &$p$, Fe  \\
\hline
\end{tabular}
\vspace{5pt}
\caption{Events analysed in examples in  
Sec.~\ref{sec:example-event}--\ref{sec:example-p-Fe}.  The columns
give (1)~reference number, (2)~experiment name (which determines the
conditions for simulations), (3)~the reported energy in units of
$10^{20}$~eV, (4) and (5)~horizontal coordinates in degrees (azimuth angle
$\phi =0$ corresponds to a particle coming from the South, $\phi
=90^\circ$ -- from the West), (6)~reported muon density at 1000~m from the
core in units of m$^{-2}$, (7)~numbers of examples in
Sec.~\ref{sec:example} where this event is used, (8)~types of primary
particles of simulated showers.
\label{tab:events}}
\end{table}
 
The artificial events F1--F8 used in examples of
Sec.~\ref{sec:example-artificial} have the same arrival directions and
experimental conditions as the corresponding events 1--8 and are listed in
Table~ \ref{tab:fake-events}.
\begin{table}
\begin{tabular}{ccccc}
\hline
No.& primary & $E_0$ & $E_{\rm obs}$ &$\rho^{\rm obs} _\mu (1000)$ \\
(1)&(2)&(3)&(4)&(5)\\
\hline
F1 & $p$       & 1.78 & 2.45 & 17.7 \\
F2 & $p$       & 1.55 & 1.84 & 10.9 \\
F3 & $\gamma $ & 1.05 & 0.94 &  1.0 \\
F4 & $p$       & 1.04 & 1.35 & 13.1 \\
F5 & $\gamma $ & 1.03 & 1.25 &  1.0 \\
F6 & $\gamma $ & 1.21 & 1.77 &  1.8 \\
F7 & Fe        & 1.65 & 2.10 &  14.9 \\
F8 & Fe        & 2.69 & 1.71 &  20.7 \\
\hline
\end{tabular}
\vspace{5pt}
\caption{Artificial events analysed in examples in
Sec.~\ref{sec:example-artificial}. The columns
give (1)~reference number, (2)~primary particle,
(3)~the original energy in units of
$10^{20}$~eV, (4)~the reconstructed energy in units of
$10^{20}$~eV,
and (5)~reconstructed muon density at 1000~m from the
core in units of m$^{-2}$.
\label{tab:fake-events}}
\end{table}

For each of the eight events, we generated libraries of simulated showers
induced by primaries required for the particular examples (see column~(8) of
Table~\ref{tab:events}). For each observed event and each primary, 1000
showers were generated.
Thrown energies
$E_0$ of the simulated showers were randomly selected
between $5\times 10^{19}$~eV and $5\times
10^{20}$~eV following the spectrum $E_0^{-1}$, as discussed in
Sec.~\ref{sec:integrals-event}. The arrival directions of the simulated
showers were the same as those of the corresponding real events. The
simulations were performed with CORSIKA~v6.204~\cite{Heck:1998vt},
choosing QGSJET~01c~\cite{Kalmykov:1997te} as high-energy and
FLUKA~2003.1b~\cite{fluka} as low-energy hadronic interaction model.
Electromagnetic showering was implemented with EGS4~\cite{Nelson:1985ec}
incorporated into CORSIKA. Possible interactions of the primary photons
with the geomagnetic field were simulated with the PRESHOWER option of
CORSIKA~\cite{Homola:2003ru}.
As suggested in
Ref.~\cite{Thin}, all simulations were performed with thinning level
$10^{-5}$, maximal weight $10^6$ for electrons and photons, and $10^4$ for
hadrons.

For each simulated shower, we determined $S(600)$ and $\rho_{\mu}(1000)$.
The plane orthogonal to the arrival direction has been divided into
concentrical rings of the width of 100~m; contributions of all particles
were averaged over these rings to reconstruct the lateral distribution
function. For the calculation of the signal density, we
used the detector response functions from
Refs.~\cite{Sakaki,YakutskGEANT}. The reconstructed energy -- the
E-observable which we compared to the experimental values -- was obtained
by making use of exactly the same procedure as used in processing the real
data; see Appendix~\ref{app:reconstruction}.

\section{Reconstruction of energy and muon density}
\label{app:reconstruction}
It is important to
process the simulated events in a way as close as possible to the
procedure with which the observables we use were obtained for the real
events. For completeness, we recall here the relevant procedures for
AGASA and Yakutsk.

\subsection{AGASA energy estimation}
\label{app:rec:AGASA}
 For the primary energy estimation, AGASA used the following procedure
(the Takeda method, Ref.~\cite{AGASAenergy}).

The lateral distribution function (LDF) of the signal is
obtained by fitting the scincillator detectors' readings by the
empirical formula~\cite{AGASA:LDF},
$$
S(r)\propto\left( \frac{r}{R_{M}} \right)^{-1.2}
\left(1+\frac{r}{R_{M}} \right)^{-(\eta-1.2)}
\left(1+\left(\frac{r}{R_{1}}\right)^{2}
\right)^{-0.6},
$$
where
$$
\eta=3.97-1.79 (\sec\theta-1), ~~~
R_{M}=91.6~{\rm m}, ~~~
R_{1}=1000~{\rm m}
$$
and $r$ denotes the distance to the shower core. Only detectors with
300~m$\le r \le$ 1000~m are used for the fit.
$S(600)$ is the value of the fitted LDF at $r=600$~m.

Due to the 10\% one-side systematic error
reported in
Ref.~\cite{AGASAenergy}, the value of $S(600)$ observed by AGASA is larger
than the real one by a factor of 1.1 which should be taken into account
for the simulated events:
$$
S^{\theta}_{\rm AGASA}(600)=1.1\,
S^{\theta}(600)
$$
The energy estimation formula is
\cite{Dai_et_al}
$$
E = 2.03 \cdot 10^{17}~{\rm eV}~
S^0_{\rm AGASA}(600),
$$
where $S^0(600)$ is the signal density for a vertical shower related to
the real density of an inclined shower
$S^{\theta}_{AGASA}(600)
$
by the attenuation
formula~\cite{AGASA:LDF}
$$
S^{\theta}_{\rm AGASA}(600) = S^0_{\rm AGASA}(600)
\exp\left(-{X_0\over\Lambda_1} \left(\sec\theta - 1\right) -
{X_0\over\Lambda_2} \left(\sec\theta - 1\right)^2\right),
$$
where
$$
X_0 = 920 ~{\rm g/cm^2}, ~~
\Lambda_1 = 500 ~{\rm
g/cm^2}, ~~ \Lambda_2 = 594 ~\rm g/cm^2
$$
The simulations should be performed for the AGASA atmospheric depth,
$
X_{\rm AGASA}\approx 960~ \rm
g/cm^2
$ (note that $X_{\rm AGASA}\ne X_0$).

\subsection{AGASA muon density estimation}
\label{app:rec:AGASA:mu}
AGASA detects the number of muon-counter hits, which is expected to be
equal to the number of muons with kinetic energy $E_{\rm kin}>0.5~{\rm
GeV}/\cos\theta_\mu $, where $\theta_\mu $ is the zenith angle of a
particular muon.

To determine the muon density at 1000 m from the core, the densities
measured by the detectors between 800 m and 1600 m from the core are
fitted by the empirical LDF~\cite{AGASA:muonLDF}
$$
\rho_\mu(r)\propto\left( \frac{r}{R_0} \right)^{-0.75} \left(
1+\frac{r}{R_0} \right)^{-\beta} \left(1+\left(\frac{r}{R_1}\right)^3
\right)^{-\delta},
$$
where
$$\beta=2.52, ~~ \delta=0.6, ~~ R_0 = 266~{\rm m}, ~~ R_1 = 800~{\rm m}.$$
The value of the fitted LDF at $r=1000$~m, $\rho _\mu (1000)$, is used as
the primary composition estimator.

\subsection{Yakutsk energy estimation}
\label{app:rec:Yakutsk}
The energy estimation in the Yakutsk
array~\cite{Yakutsk:energy,Yakutsk:energy-add} is quite similar to
that of AGASA. The value of $ S(600) $ is determined by fitting the
scincillator detectors' readings by the following empirical formula,
$$
S(r)\propto \left( \frac{r}{R_{M}} \right)^{-a}
\left(1+\frac{r}{R_{M}} \right)^{a-b}
\left(1+\left(\frac{r}{R_{1}}\right) \right)^{-g},
$$
where~\footnote{The actual procedure is more
involved\cite{MP-message}. It takes into account the atmospheric
conditions at the arrival time of a given event, that in particular
results in an event dependence of $R_M$ which varies usually within
$60~{\rm m}<R_M<80$~m. The detector responses then are adjusted in
order to get the same energy estimate with $R_{M}=68.0~{\rm m}$, that
allows to treat the data uniformly. Since this procedure adopted by
Yakutsk collaboration is unambigiuos, our simple method is justified.}
$$
R_{M}=68.0~{\rm m}, ~~
R_{1}=2000~{\rm m}, ~~
a=1.3,              ~~
b=3.24-2.6 (1-\cos\theta), ~~
g=3.5.
$$ The attenuation formula is
$$
S(600) = S^0(600) \left(
(1-\beta) \exp \left(-\frac{X_0}{\Lambda_E} \left(\sec\theta - 1\right)
\right) +\beta \exp \left(-\frac{X_0}{\Lambda_M} \left(\sec\theta - 1
\right) \right) \right).
$$
The energy  conversion formula is
\[
E = 4.6 \cdot 10^{17}~{\rm eV}~ \left[S^0(600)\right]^{0.98},
\]
where
$$
X_0 = 1020 ~{\rm g/cm^2}, ~~
\Lambda_E = 250 ~{\rm
g/cm^2},~~\Lambda_M = 2500 ~{\rm g/cm^2}, ~~
\beta=0.39 \cdot \left[ S^0(600)\right]^{-0.12}.
$$
The simulations are to be performed for the Yakutsk atmospheric depth,
$
X_{\rm Yakutsk}=X_0\approx 1020~ \rm
g/cm^2
$.

\subsection{Yakutsk muon density estimation}
\label{app:rec:Yakutsk:mu}
For the two events we use, the Yakutsk muon detectors counted muons with
$E_{\rm kin}>1.0~{\rm GeV}/\cos\theta_\mu $.

To apply one and the same procedure to both AGASA and Yakutsk results, we use
the muon density at 1000 m as the parameter to be compared with
simulations. The Yakutsk muon detectors have larger area and
significantly higher saturation threshold than AGASA's, so the detector
readings in a wider region, 400~m to 2000~m, are used in the LDF fit by
the following empirical formula~\cite{Yakutsk-mu-LDF,Yakutsk:energy}
$$
\rho_\mu(r)\propto \left( \frac{r}{R_0} \right)^{-0.75} \left(
1+\frac{r}{R_0} \right)^{0.75-b_\mu}
\left(1+\frac{r}{R_1}\right)^{-g_\mu},
$$
where $
R_0 = 280~{\rm
m}$, $
R_1 = 2000~{\rm m}
$, $g_\mu =8$ and $b_\mu $, together with the overall proportionality
coefficient, is a free fitting parameter.

\end{document}